\documentclass[a4paper,10pt]{article}
\usepackage[utf8]{inputenc}

\usepackage[top=2.0cm, bottom=2.0cm, left=2.5cm, right=2.5cm]{geometry}
\usepackage{multicol}

\usepackage{amsmath,amssymb}
\usepackage{graphicx}
\usepackage{textcomp}
\usepackage{subfigure}

\title{Suppression of the Rayleigh-Plateau instability on a vertical fibre coated with wormlike micelle solutions\footnote{Accepted in Soft Matter. DOI:10.1039/C3SM27940E}}
\author{F. Boulogne\textsuperscript{1,2}, M.A.~Fardin\textsuperscript{3,4}, S. Lerouge\textsuperscript{3,4},\\ L. Pauchard\textsuperscript{1}, F. Giorgiutti-Dauphin\'e\textsuperscript{1}}

\begin{document}

\maketitle
\emph{\textsuperscript{1} UPMC Univ Paris 06, Univ Paris-Sud, CNRS, F-91405. Lab FAST, Bat 502, Campus Univ, Orsay, F-91405, France.\\
\textsuperscript{2} boulogne@fast.u-psud.fr\\
\textsuperscript{3} Laboratoire Mati\`ere et Syst\`emes Complexes, CNRS UMR 7057, Universit\'e Paris Diderot, 10 rue Alice Domon et L\'eonie Duquet, 75205 Paris C\'edex 13, France.\\
\textsuperscript{4} The Academy of Bradylogists
}

\begin{abstract}
We report on the Rayleigh-Plateau instability in films of giant micelles  solutions coating a vertical fibre. 
We observe that the dynamics of thin films coating the fibre could be very different from the  Newtonian or standard Non-Newtonian cases. 
By varying the concentration of the components of the solutions and depending on the film thickness, we show for the first time that the Rayleigh-Plateau instability can be stabilized using surfactant solutions. 
Using global rheology and optical visualisations, we show that the development of shear-induced structures is required to stabilize the micellar film along the fibre. 
Assuming that the viscoelastic properties of the shear-induced state can be described by a simple model, we suggest that, in addition to the presence of shear-induced structures, the latter must have an elastic modulus greater than a critical value evaluated from a linear stability analysis. 
Finally, our analysis provides a way of estimating the bulk elasticity of the shear-induced state.
\end{abstract}

\section{Introduction}

When a liquid film is coating a fibre, it undergoes spatial thickness variations. 
The mechanism of this instability has been understood by Lord Rayleigh in the 19th century\cite{Rayleigh1878}. 
Due to the energetic cost of free surfaces (surface tension), the liquid tends to minimize its surface area by breaking a cylinder into a serie of regularly spaced droplets.
This instability can be observed on free falling jets and it underlies many natural phenomena, like the dew drops on coweb \cite{Boys1959}. 
The coating of fibres has been extensively studied by Qu\'er\'e \textit{et al.} in the situation where the fibre is drawn out of a bath \cite{Quere1999}.
They compared their measurements with theoretical predictions from Landau, Levich and Derjaguin on the film thickness \cite{Landau1942,Derjaguin1943}.

In the case of a Newtonian fluid flowing down a vertical fibre, flow regimes have been depicted as a function of physical parameters. 
For high flow rates and/or fibre radii, the flow is dominated by inertia and the nature of the instability is convective \cite{Duprat2007}. 
On the contrary, when the fibre size is smaller than the capillary length and for creeping flow (low Reynolds number), the instability is absolute and the physics is dominated by surface tension \cite{Duprat2007}.
In such absolute regimes, we recently studied  the flow of polymer solutions \cite{Boulogne2012}. 
We investigated experimentally the influence of two non-Newtonian properties, shear-thinning effect and first normal stress difference, on the growth rate of the instability and on the morphology of the drops. The pattern is globally the same as in the case of Newtonian fluids: the first normal stress difference just tends to slightly decrease the growth rate and to smooth the drop shape.

Beside polymers, a broad variety of molecules assemble to form non-Newtonian fluids. 
For instance, it is well-known that above a Critical Micellar Concentration (CMC), surfactant molecules can self-assemble to form aggregates called micelles. 
The size and the shape of the micelles depend on the structure of the surfactant molecule, on the surfactant concentration and on the presence of additives like simple or organic salts \cite{Israelachvili2011}. 
In some range of parameters, the micelles are giant worms often called "living polymers" because thay can entangle like polymers but they can also break and fuse continuously.
Solutions of worm-like micelles present remarkable rheological properties such as shear-thickening or shear-banding effects, extensively studied by theoreticians and experimentalists \cite{Berret2006,Cates2006}. 
These non-linear properties are often associated with the development of out of equilibrium structures induced by the shear flow.
These shear-induced structures present strong viscoelastic properties leading to specific behaviors like the oscillation of a falling sphere \cite{Jayaraman2003} or the incomplete retraction of a filament after a pinch-off \cite{Smolka2003}.
These examples highlight the consequence of the viscoelastic properties of the flow-induced phases.

In this paper, we investigate the flow, at low Reynolds number, along a vertical fibre of semi-dilute giant micelles solutions. By varying the concentration of the chemical compounds of the solutions and the thickness of the film along the fibre, we observe different morphologies for the micellar film. In some conditions, the film exhibits the expected Rayleigh-Plateau instability, as it has been observed so far, in Newtonian fluids or viscoelastic polymer solutions while for other conditions, remarkably, the micellar film remains stable. Using global rheology and optical visualisations, we demonstrate that the stabilisation process along the fibre is connected with the development, above a characteristic shear stress, of shear-induced structures. We also show that the development of shear-induced structures is necessary but not sufficient to stabilize the micellar film. Using a simple viscoelastic model (Kelvin-Voigt) to describe the viscoelastic properties of the shear-induced state, we suggest that the 
elastic modulus characterizing the shear-induced structures has to be greater than a critical value provided by the linear stability analysis.

The paper is organized as follows.
In Section \ref{sec:exp_setup}, we present the experimental setup and the chemical system. 
In Section \ref{sec:flow_regime}, we report different flow regimes depending on surfactant and/or salt concentrations and film thicknesses.
In Section \ref{sec:rheo}, we present the rheological behavior of our micellar systems and we visualize the material under shear to confirm the presence of shear-induced structures.
The next two Sections are devoted to the conditions required to stabilize the micellar film on the fibre.  
Section \ref{sec:SIS} focus on the stress applied to the film by gravity and its relation with the rheological behaviour of the solution and 
Section \ref{sec:elasticity} studies the role of the shear-induced film elasticity on the Rayleigh-Plateau instability.

\section{Experimental details}\label{sec:exp_setup}

\subsection{Experimental setups}

\begin{figure}
\centering
\subfigure[]{\includegraphics[scale=0.90]{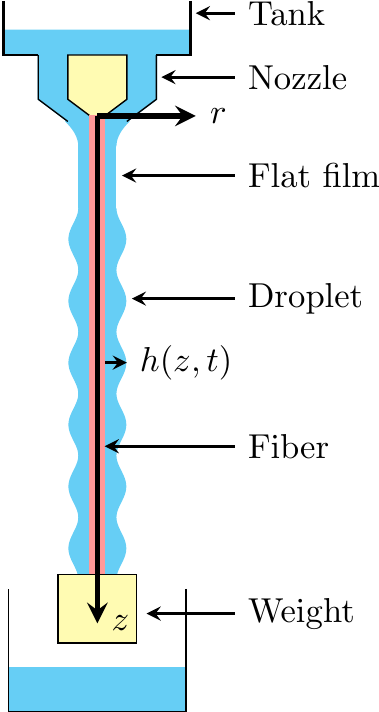}}
\subfigure[]{\includegraphics[scale=0.33]{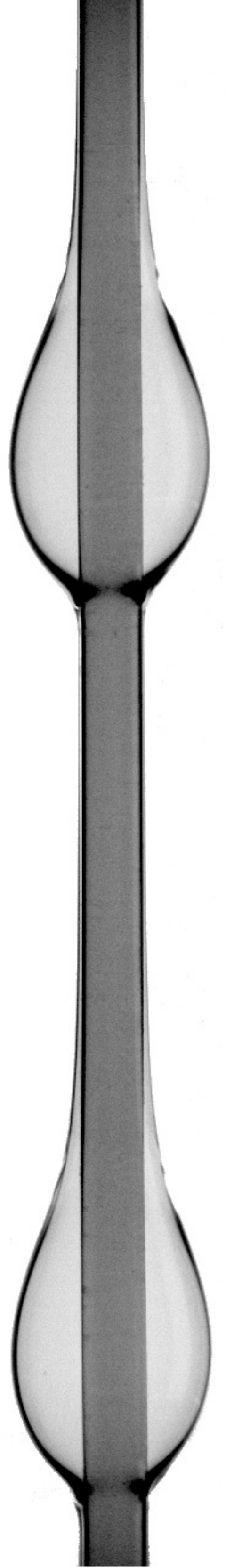}}
\subfigure[]{\includegraphics[scale=0.33]{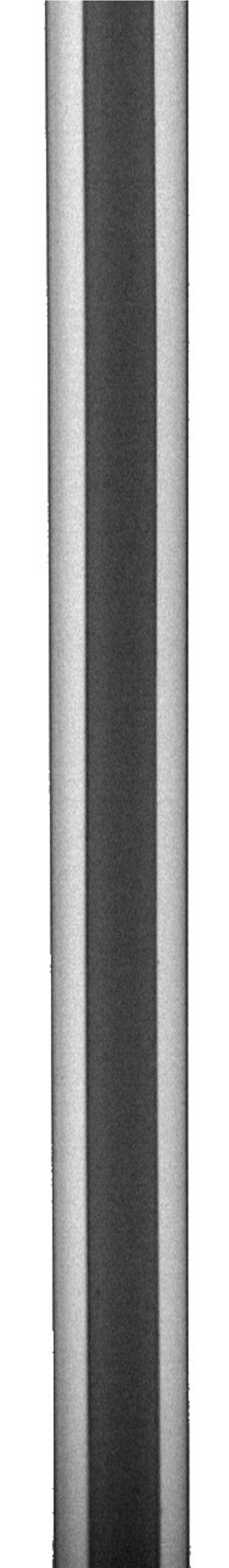}}
\subfigure[]{\includegraphics[scale=0.33]{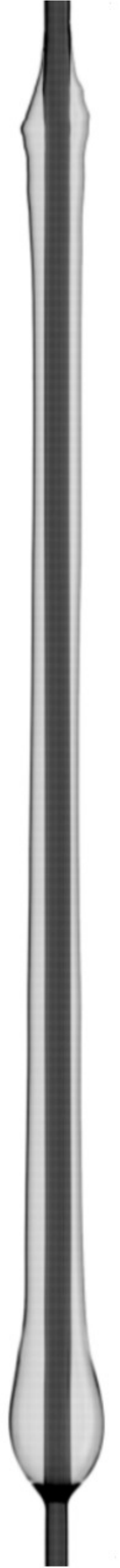}}

 \caption{(a) Notations and schematic view of the vertical fibre and the injection device. 
(b-d) Pictures illustrating the different morphologies of the film flowing along the fibre: 
(b) Unstable morphology which corresponds to drops sliding on a quasi static liquid film. 
(c) Stable film. 
(d) Film morphology corresponding to a gel-like block moving down the fibre.  
The diameter of the fibre is $0.56$~mm.}\label{fig:setup}
\end{figure}

The experimental setup used for the study of the Rayleigh-Plateau instability is depicted in Figure \ref{fig:setup}(a). 
An upper tank, of diameter $14$ cm, is connected to a valve. 
This valve is composed of two axisymmetric cones. 
The adjustment of the gap between these cones controls the flow rate \textit{i.e.} the film thickness on the fibre.
Guided by a nozzle, the liquid flows along a vertical nylon fibre. 
The fibre is  about $60$ cm long and has a radius $R=0.28$ mm.
Vertical position and centering of the nozzle are crucial to ensure an axisymmetric coating. 
This position is adjusted using an optical support with a precision of 2.4 arc sec (Model U200-A2K, Newport).
Observations are done with a telecentric lens and a high speed camera (about $500$ images/s) delivering a resolution of $0.024$ mm/pixel (Fig.~\ref{fig:setup}(b-c)).

The rheological properties of the samples have been characterized using a stress-controlled rheometer (Anton Paar, MCR 501) at $20.00\pm0.01^\circ$C. 
The temperature was adjusted with a Peltier plate. We used a cone and plate geometry (radius $49.988$mm, angle $0.484^\circ$). 
A solvent trap was used to limit evaporation during the measurements. 
Rheological protocol consists in applying stress sweep, the stress value being imposed during 1000~s at each step. The flow curve is obtained by averaging the shear rate response over the last 100~s of each step.

The global rheological measurements have been complemented by rheo-optical observations of the flow performed in a Taylor-Couette (TC) flow geometry 
(gap: $1.13$ mm, length: $4$ cm), adapted to the shaft of a stress-controlled rheometer (Anton Paar MCR 301). 
The inner cylinder is rotating while the outer transparent cylinder is fixed in the laboratory frame. 
Direct visualizations were made in the plane velocity gradient/vorticity (\textit{i.e.} the plane made by the radial direction and the cylinders axis), the gap of the TC cell being illuminated with white light. 
Images of the gap were collected using a CCD camera. 
The rheological signal was recorded simultaneously, allowing direct correlation with optical visualizations \cite{Lerouge2008,Fardin2009}. 
Note that similar rheograms were obtained with TC and cone and plate geometries.

\subsection{Materials}

We focus on aqueous micellar solutions made of cetyltrimethylammonium bromide (CTAB) and sodium salicylate (NaSal) purchased from Sigma-Aldrich and used without further purification.
NaSal is an organic salt, sometimes designated as a co-surfactant since it participates to the micellar structure by taking place in-between the polar head groups of surfactant molecules \cite{Rao1987}.
The addition of this salt also screens repulsive electrostatic interactions between CTA$^+$ groups and it reduces the spontaneous curvature of the system \cite{Lin1994}.
As a result, the micellar uniaxial growth is promoted because of the large energetic cost of hemispherical endcaps. 
All the solutions investigated range in the semi-dilute regime. For the range of concentration chosen here, the micelles are locally cylindrical and slightly entangled~\cite{Berret2006}.

To prepare chemical solutions, we mixed in pure water (milliQ quality, resistivity: $18$ M$\Omega$.cm) the weight percentages of NaSal and CTAB until a complete dissolution.
Samples were stored at rest at $23^\circ$C in darkness to avoid any degradation.
The solutions were allowed to reach equilibrium for at least three days in these conditions before experiments were performed.
We carefully checked the homogeneity of solutions before any measurement.
All measurements along the fibre were carried out at $T=20.0\pm0.5^\circ$C.

The equilibrium surface tension of the solutions studied in this paper is $36.0\pm0.5$ mN/m (Wilhelmy plate method with a Kr\"uss tensiometer), a value consistent with measurements reported in the literature for solutions above the critical micellar concentration \cite{Cooper-White2002}.

\section{Morphologies of the liquid film along a vertical fibre}\label{sec:flow_regime}

We focused on low Reynolds number flows ($Re \lesssim 1$) dominated by surface tension ($R<l_c = \sqrt{\frac{\gamma}{\rho g}} = 1.9$ mm), for which the flow is absolutely unstable \cite{Duprat2007}. 
We studied the flow along the vertical fibre of a large set of wormlike micelles solutions. 
Depending at once on the surfactant concentration [CTAB], the salt concentration [NaSal] and the film thickness  $h$, we observe three possible morphologies of the film along the fibre as displayed in Fig.~\ref{fig:setup}(b-d): 
\begin{itemize}
    \item The film can be unstable with respect to the Rayleigh-Plateau instability. The associated morphology corresponds to drops sliding on a quasi-static liquid film (Fig.~\ref{fig:setup}(b)).
    \item The film can be stable and in this case, the associated morphology corresponds to a flat film (Fig.~\ref{fig:setup}(c)).
    \item The film can be subjected to some kind of fracture leading to separate blocks moving down the fiber  (Fig.~\ref{fig:setup}(d)). The morphology is essentially flat but the film is discontinuous.
\end{itemize}

\begin{figure*}
\begin{center}
\includegraphics[scale=1]{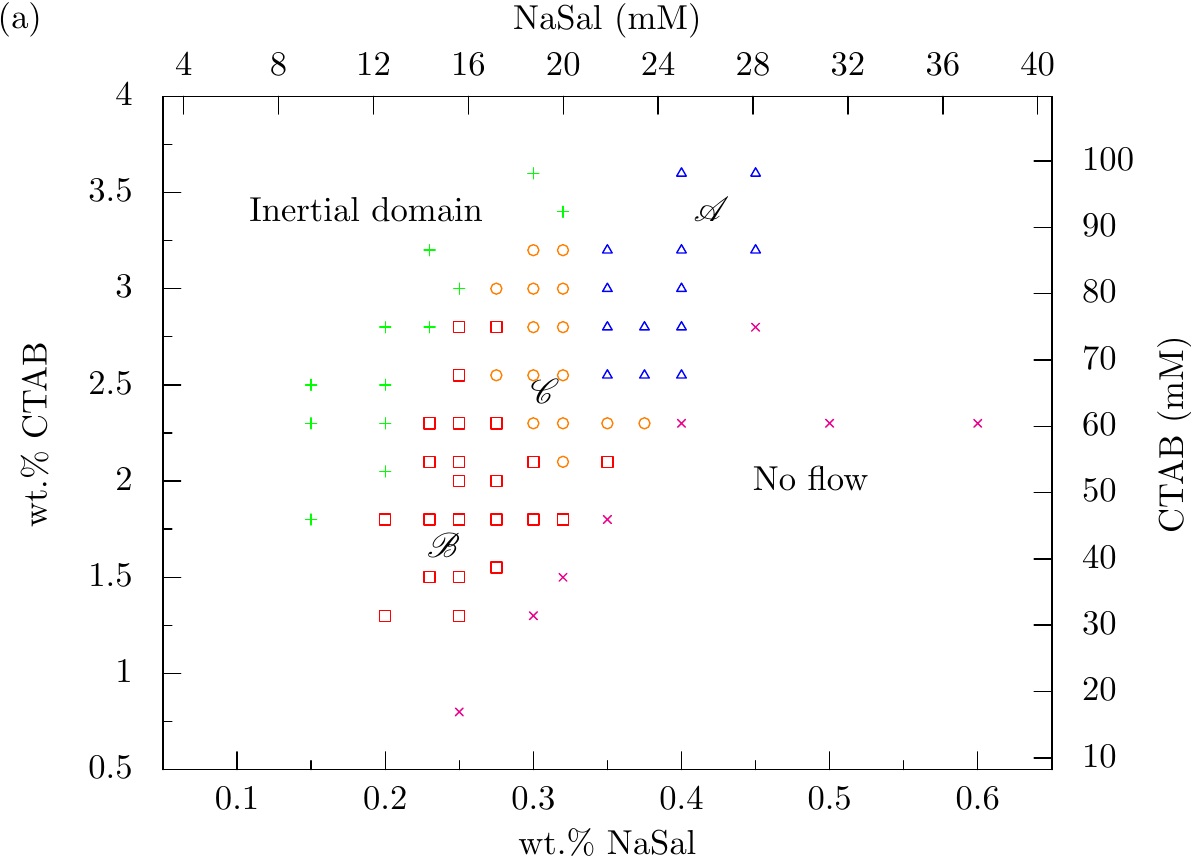}
\includegraphics[scale=1]{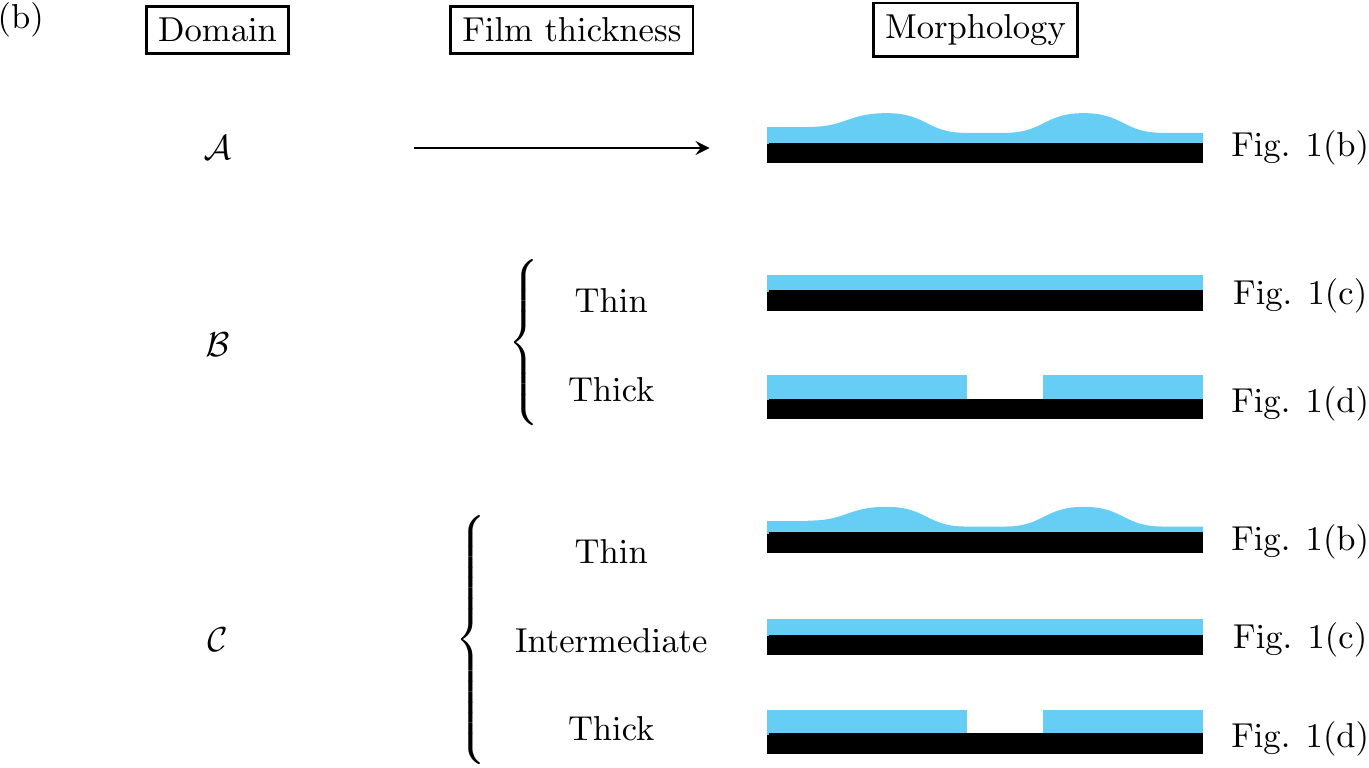}
\end{center}
 \caption{(a): Flow-phase diagram showing domains depending on surfactant (CTAB) and cosurfactant (NaSal) concentrations. 
 Five domains are identified: Inertial (green $+$), "No flow" (magenta $\times$), $\cal{A}$ (blue $\triangle$), $\cal{B}$ (red $\Box$) and $\cal{C}$ (orange $\circ$).
 (b): Schematic representations of the different sequences of film morphologies  as a function of the film thickness associated with the different domains $\cal{A}$, $\cal{B}$ and $\cal{C}$ in the flow-phase diagram.
}\label{fig:diagram}
\end{figure*}

In order to illustrate the effect of the various parameters on the flow along the fibre, we construct a flow-phase diagram (see Fig.~\ref{fig:diagram}a) in the plane ([CTAB], [NaSal]), allowing to identify different domains.

At low NaSal concentrations, the apparent viscosity is low ($\lesssim 50$ mPa.s) which corresponds to inertial regimes \cite{Duprat2007}. 
At high NaSal concentrations, due to a high zero shear viscosity ($>100$ Pa.s), the fluid hardly or could not flow though our experimental device ("No flow" domain). 
These two domains are excluded from the present study which, focuses on the intermediate range of NaSal concentrations, typically between $0.2$ and $0.5$ wt.\%. 
In this intermediate range, three domains, denoted by $\cal{A}$, $\cal{B}$ and $\cal{C}$, can be distinguished according to the sequence of morphologies observed as a function of the film thickness.

\begin{itemize}
 \item  Domain $\cal{A}$
 
In this domain, the micellar solutions behave as usual Newtonian fluids \cite{Duprat2007} or polymer solutions \cite{Boulogne2012} for all the film thicknesses accessible with our experiment: the film is first flat and then, the Rayleigh-Plateau instability develops (Fig.~\ref{fig:setup}(b)). 

 \item Domain $\cal{B}$
 
In this domain, regardless of the film thickness, the micellar solutions do not exhibit the Rayleigh-Plateau instability.
For sufficiently small thicknesses (\textit{i.e.} low flow rates), the film remains flat and continuous over the whole fibre extent (about $60$ cm). 
From this stable morphology, an increase of the film thickness leads to a breaking of the liquid thread into blocks of a few centimeters long, sliding along the fibre (Fig.~\ref{fig:setup} (d)). Note that, visually the blocks seem to have a gel-like character with a large elasticity, in contrast to the solutions at rest.
Similar viscoelastic threads were observed in the case of pinch-off and filament retraction in extensional flow of wormlike micelles \cite{Smolka2003,Bhardwaj2007}.

 \item Domain $\cal{C}$
 
     In between the two domains $\cal{A}$ and $\cal{B}$, the domain $\cal{C}$ is characterized by the existence of the three morphologies (see Fig.~\ref{fig:setup}(b-d)). 
For small thicknesses, the films are destabilized by the Rayleigh-Plateau instability with patterns  similar to those described in domain $\cal{A}$. 
As its thickness is increased, the film adopts a stable morphology corresponding to a flat continuous liquid thread. 
Finally, for sufficiently large thicknesses, the flat and continuous liquid thread breaks into gel-like blocks as described in domain $\cal{B}$.

\end{itemize}

\begin{figure}
\begin{center}
\includegraphics[scale=0.6]{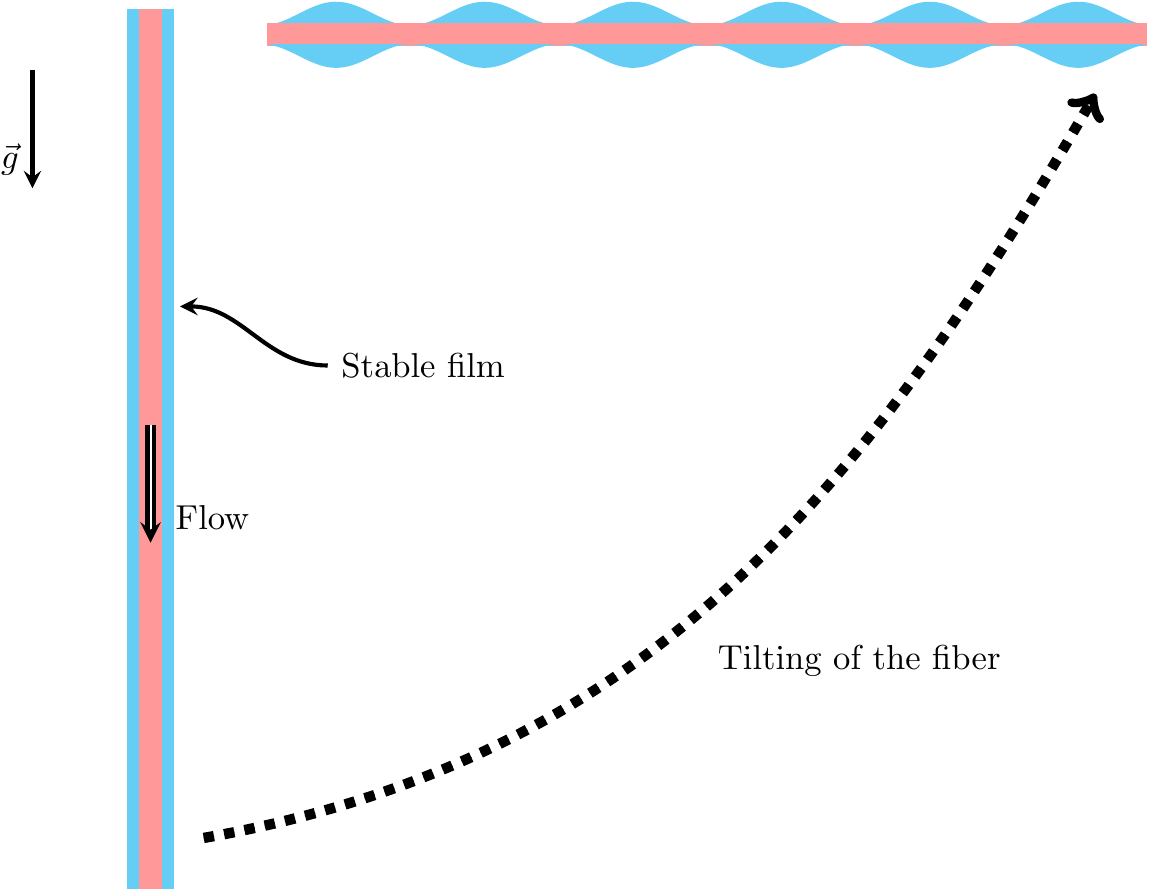}
\end{center}
 \caption{From a vertical fibre where a stable film is flowing down, we tilt the fibre horizontaly leading to the development of the Rayleigh-Plateau instability.}\label{fig:tilt}
\end{figure}

The different sequences of morphologies described above and observed respectively in the domains $\cal{A}$, $\cal{B}$ and $\cal{C}$ are schematized in Fig.~\ref{fig:diagram}(b).

When a stable morphology is observed, we have noticed that if we tilt the fibre horizontally, after a few
seconds, drops can appear (figure \ref{fig:tilt}). This observation suggests that 
the flow under the action of gravity is necessary to stabilize the film.
Particularly, this shows that the flow properties of the micellar solutions are involved.

Note that this effect is different from the stabilizing process depicted by Qu\'er\'e \cite{Quere1990} with Newtonian fluids. 
In his experiments, stable films flowing down a vertical fibre were obtained by saturating the Rayleigh-Plateau instability thanks to  the advection of the flows.
This saturation (\textit{i.e.} flat film) occurs if the film thickness is smaller than $h_c = \frac{R^3}{l_c^2}$.
In our experiment, $h_c=6 \mu$m while $h>200 \mu$m, and the flow is absolutely unstable .

\section{Rheo-optical properties under simple shear flow}\label{sec:rheo}
\begin{figure}
\begin{center}
\includegraphics{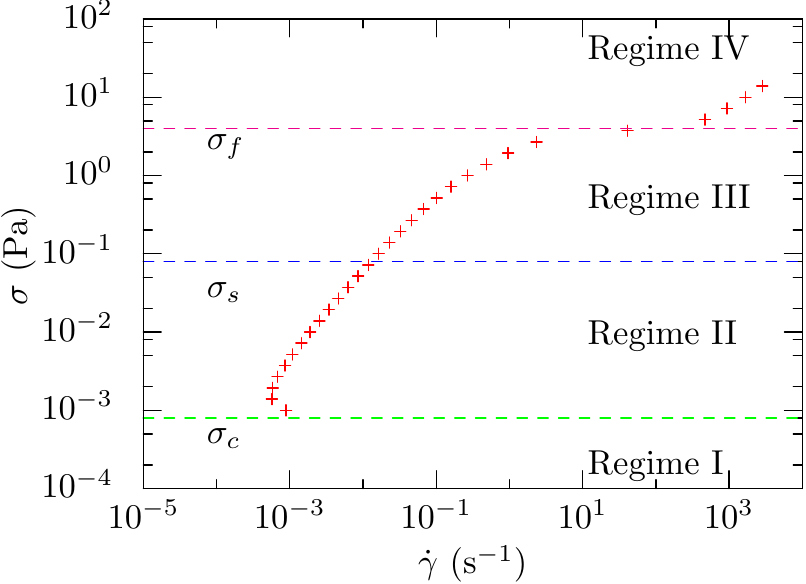}\\
\includegraphics{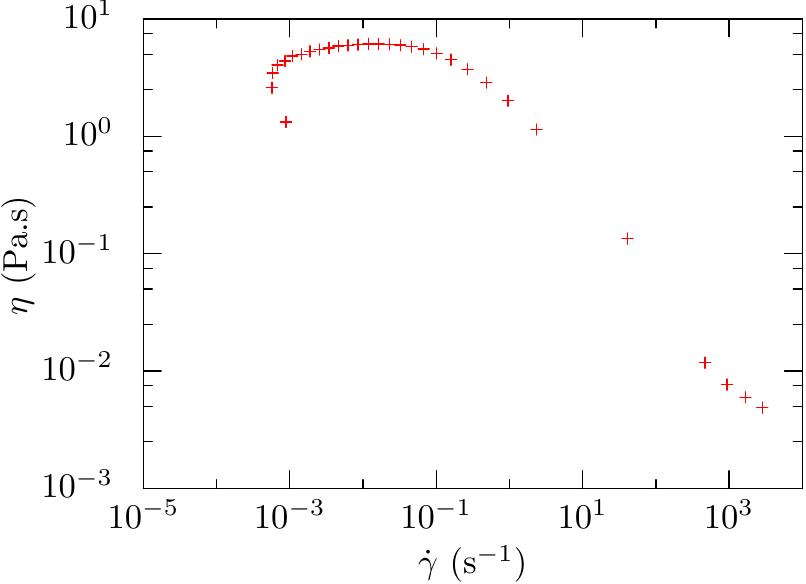}

\end{center}
 \caption{(a): Shear stress $\sigma$ and (b): apparent viscosity $\eta$ as a function of the shear rate $\dot\gamma$ for a solution of CTAB ($2.55$ wt\%)  and  NaSal ($0.32$wt.\%). 
 The temperature is T=$20.00\pm0.01^\circ$C. 
 The rheological protocol consists in stress sweep experiment: at each step, the shear stress is  kept fixed during $1000$~s. 
 The resulting shear rate $\dot\gamma(t)$ reaches a stationnary value after a time period that depends on the applied shear stress. 
 The flow curve is obtained by averaging the measured shear rate over the last $100$~s of each step. 
 $\sigma_c$, $\sigma_s$ and $\sigma_f$ denotes characteristic shear stresses that delimit different regimes in the flow curve.}\label{fig:rheology}
\end{figure}

\begin{figure*}
\begin{center}
 \includegraphics{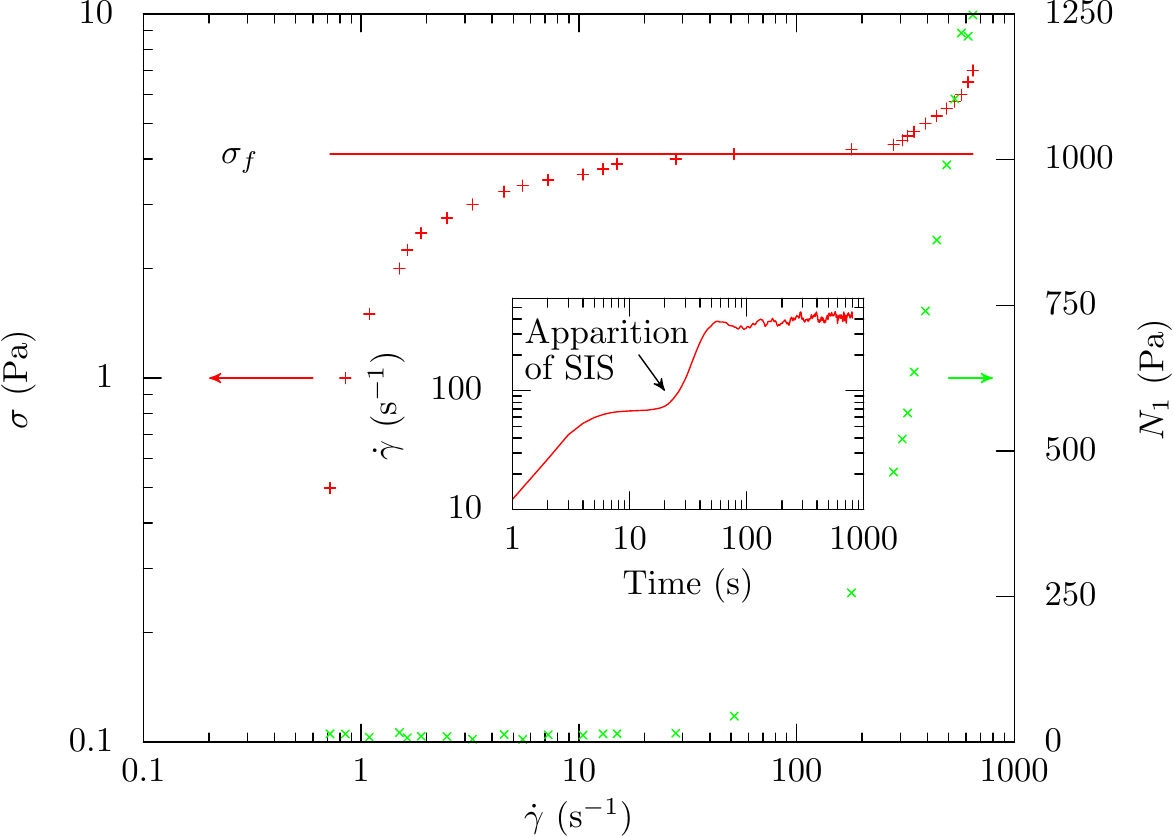}
\end{center}
 \caption{Determination of the stress plateau value ($\sigma_f$). Stress $\sigma$ (red $+$) and first normal stress difference $N_1$ (green $\times$) versus shear rate for $2.55$wt.\% CTAB and $0.32$wt.\% NaSal (T=$20.00\pm0.01^\circ$C). 
 At each step, the shear stress is imposed during $1000$~s and the measured shear rate results from the average over the last $100$~s of each step. 
Inset: Time evolution of the shear rate $\dot\gamma$ for a fresh sample at $\sigma=4.5$ Pa. The black arrow indicates the signature linked to the apparition of SIS.}\label{fig:rheology_plateau}
\end{figure*}

\begin{figure}
\centering\includegraphics[scale=0.3]{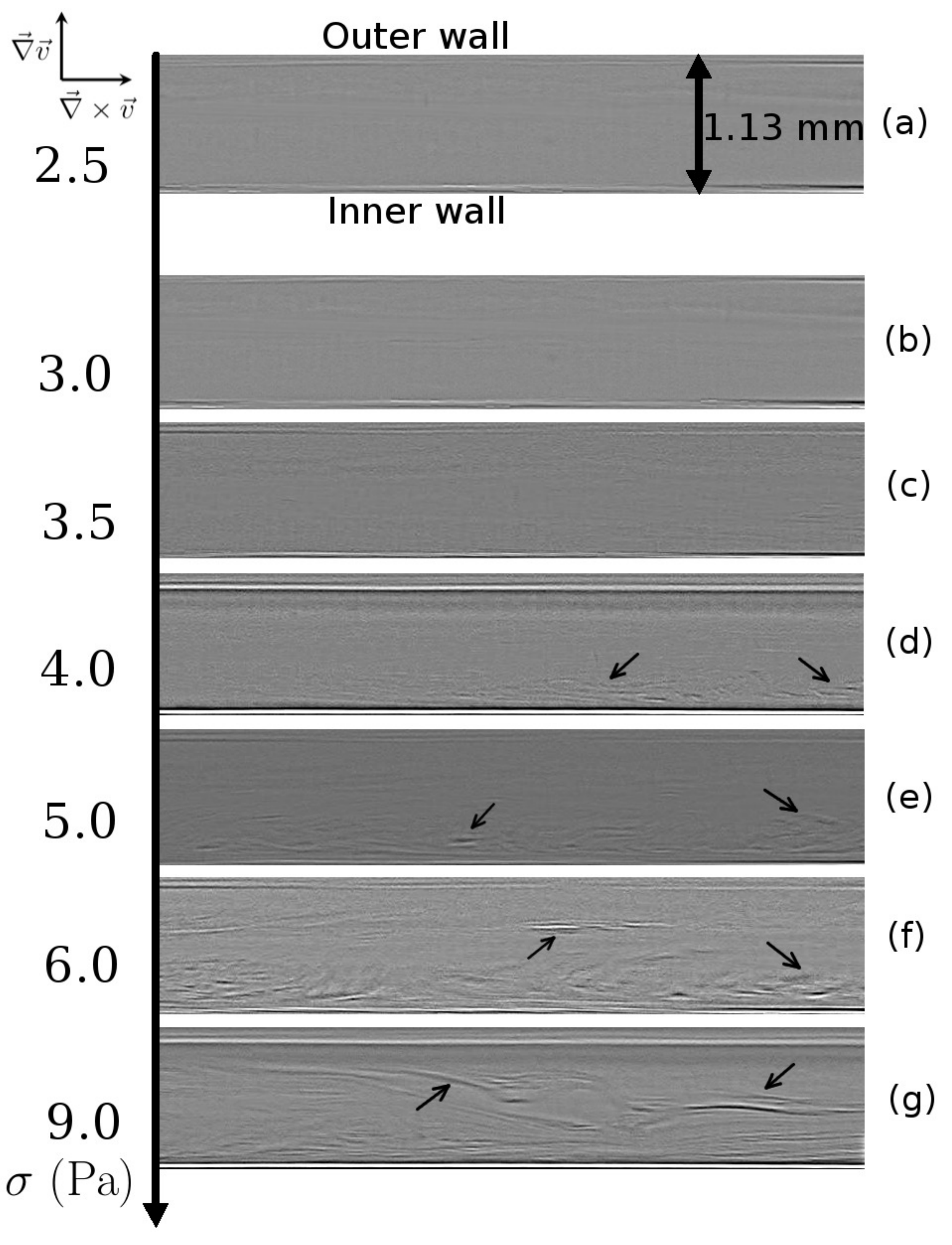}
\caption{Views of the gap of a Couette cell in the plane velocity gradient/vorticity (noted $(\vec{\nabla} \vec{v}, \vec{\nabla}\times \vec{v})$) for different applied shear stresses. 
 Each image is taken after a time $t\sim 1000$ s. 
 The solution is composed of $2.55$wt.\% CTAB and $0.32$wt.\% NaSal. 
 The top and bottom sides correspond to the outer and inner wall respectively. 
 For $\sigma>\sigma_f=4$ Pa, structures appear near the inner wall. 
 Arrows indicate some turbidity fluctuations indicative of SIS with micronic characteristic length scale.}\label{fig:sis}
\end{figure}

In contrast to the free jets dominated by extensionnal flow~\cite{Clasen2006,Eggers2008,Bhat2010}, the flow along the fibre is dominated by shear. 
We thus performed rheological experiments under simple shear flow  in order to characterize the rheological properties of the micellar solutions used in this study. 
The measurement of the flow properties is combined with direct optical visualisations of the samples illuminated with white light in order to detect possible changes in the structure of the micellar fluid. 
In the different domains $\cal{A}$, $\cal{B}$ and $\cal{C}$ identified in the flow-phase diagram, we observe the same type of flow behaviour. 
Figure~\ref{fig:rheology}(a) displays a representative flow curve gathered from a sample belonging to domain $\cal{C}$ and obtained in stress-controlled mode.
The choice of this mode has been motivated by the fact that the maximum stress in the liquid flowing down the fibre can be easily estimated (See Section \ref{sec:SIS}).
The response of the samples to  simple shear flow is highly nonlinear, with successive transitions, from shear-thickening to different degrees of shear-thinning. Such complex evolution of the shear stress as a function of the shear rate has already been observed for dilute micellar solutions, well-known to exhibit a shear-thickening transition \cite{Hu1998,Hu1998a}. 
Following Refs. \cite{Hu1998,Hu1998a}, we divide the flow curve into four distinct regimes, defined by ``critical'' shear stresses, denoted $\sigma_c$, $\sigma_s$ and $\sigma_f$.
\begin{itemize}
\item Regime I ($\sigma<\sigma_c$)

This regime corresponds to the primary response of the micellar solution to exceedingly low shear stresses. 
For dilute systems, we expect a linear increase of the shear stress as a function of the shear rate, associated with a Newtonian behaviour while for semi-dilute systems, a slight decrease of the viscosity as a function of the shear rate is usually observed, due to partial alignment of the micelles by the flow~\cite{Hartmann1997, Hu1998}. 
Unfortunately, reaching this regime requires shear stresses that cannot be imposed by the rheometer, what leads to the lack of experimental data points in Fig.~\ref{fig:rheology}. 
Note that in these conditions, $\sigma_c$ is only roughly defined.

\item Regime II ($\sigma_c<\sigma<\sigma_s$)

This regime is characterized by a shear-thickening transition, the apparent viscosity of the material increasing with the shear rate (see Fig.~\ref{fig:rheology}). 
This transition is associated with a re-entrant behavior of the flow curve since the shear rate first decreases and then increases with the shear stress in this regime. 
Due to the re-entrant character of the transition, this regime can only be observed at imposed stress. 
The upper boundary of this regime is noted $\sigma_s$ and corresponds to the shear stress for which the apparent viscosity reaches a maximum. 
Images of the sample in the shear-thickening regime do not exihibit any changes compared to the situation at rest. 
The gap of the TC cell appears homogeneous suggesting that the changes in the structure of the fluid associated with the shear-thickening transition, occur at sub-micronic scale.

\item Regime III ($\sigma_s<\sigma<\sigma_f$)

In this range of applied stresses, the flow becomes shear-thinning and direct visualisations show that the gap of the TC cell remains homogeneous (see Fig.~\ref{fig:sis} (a-c)).

\item Regime IV ($\sigma>\sigma_f$)

This regime is characterized by a strong degree of shear-thinning associated with the existence of a stress plateau, the onset of which is denoted by a ``critical'' shear stress  $\sigma_f$. 
The stress plateau generates a jump in the shear rate resulting in a sharp decrease of the apparent viscosity.  
The stress $\sigma_f$ is determined by increasing the shear stress step by step with an appropriate sampling as illustrated in Fig.~\ref{fig:rheology_plateau}. 
The plateau is characterized by a large jump in the shear rate at imposed stress, or equivalently a strong increase of the shear rate versus time at fixed stress (see inset in Fig.~\ref{fig:rheology_plateau}) and a large increase of the first normal stress difference $N_1$, indicative of nonlinear viscolelastic properties of the material under shear. 
Note that below or above $\sigma_f$, variations of the shear rate are lower than $10$\% over the duration of the measure ($1000$ s). For the sample in Fig.~\ref{fig:rheology_plateau}, we estimate $\sigma_f = 4.1\pm 0.3$ Pa. 
In this regime, direct visualisations in the TC geometry reveal a drastic change in the structure of the fluid (see Fig.\ref{fig:sis} (d-e)). A fraction of the sample becomes slightly turbid leading to variations in the refraction index observable with visible light. These observations suggest that, for $\sigma>\sigma_f$,  a new phase characterized by a length scale in the order of a few microns nucleates in the gap of the TC device~\cite{schubert2004}. We observed that the nucleation occurs from the edges of the cylinders and progressively extends along the inner cylinder, the growth being favoured at the bottom of the cell which has a conic shape. At fixed stress, the turbidity fluctuations observed on the pictures evolve in time with a complex dynamics.   
The proportion of the induced phase increases with the imposed shear stress, and for sufficiently high shear stresses, the induced phase invades the whole gap (see Fig.\ref{fig:sis} (f-g)).
This scenario is reminiscent of the shear-banding transition even if there is no clear evidence for an organisation into two bands separated by a well-defined interface~\cite{Lerouge2010}. By monitoring the shear-rate as a function of time, we noticed that the appearance of the induced phase is correlated with the presence of a kink in the $\dot\gamma(t)$ curve which appears sooner as the stress increases.
 This kink is also observed in the cone-plate geometry for $\sigma>\sigma_f$ (See inset in Figure \ref{fig:rheology_plateau}) and is then the signature for the appearance of the induced phase.
\end{itemize}

Beyond this purely phenomenological description, our observations deserve further comments. 
As mentioned above, similar complex succession of rheological transitions in the flow curve have already been observed in dilute micellar solutions of TTAA/NaSal~\cite{Hu1998,Hu1998a}. 
However, the direct observations reported by the authors largely differ from the present study: Indeed, in the shear-thickening regime (regime II), the fluid phase of regime I was found to coexist with a viscous phase of gel-like shear-induced structures while in regime III, the induced gel-like phase invaded the whole gap of the flow geometry. 
Furthermore, the rheological response in regime III was also different, the apparent viscosity being constant, while in our case the system is shear-thinning. 
This behaviour was ascribed to the existence of a plug flow with slip at the walls. Finally, above $\sigma_f$, fracture of the gel-like phase followed by flow instabilities were observed.

Here we do not observe any phase coexistence in regime II. However, this does not rule out the fact that the shear-thickening transition is associated with a change of the structure of the system but rather suggests that this change of the structure occurs at sub-micronic scale which prevents direct visualisation. Our situation could correspond to the minimal scenario for the shear-thickening transition in micellar systems, namely the uniaxial growth of the micelles~\cite{Berret2001,Protzl1997,Oda2000,Lerouge2010}. 
On the other hand, we clearly observe a phase coexistence in regime IV, above $\sigma_f$, the induced phase being characterized by a large first-normal stress difference. 
Note that, at equilibrium, the samples under investigation are only weakly entangled: their concentration is associated with the very beginning of the semi-dilute  regime, in a ``transition''  range where the rheological behaviour is between pure shear-thickening of dilute rod-like micellar systems, and pure shear-banding of strongly entangled wormlike micelles~\cite{Hartmann1997a,Lerouge2010}. 
In this context, the shear-banding-like  behaviour observed in the present study is likely to correspond to a transition from the state induced  during the shear-thickening regime (regime II) towards another induced ``phase'' structured at larger scale (regime IV).  

Note that in the literature on wormlike micelles, the shear-induced state associated with the shear-thickening transition was  called SIS for shear-induced structure~\cite{Ohlendorf1986}, or SIP for shear-induced phase~\cite{Boltenhagen1997} interchangeably. Recent studies~\cite{Vasudevan2010} have also used FISP to denote irreversible ``flow-induced structured phase" obtained in extensional flow. However, in the present paper, we make no distinction between the expressions ``shear-induced structures" and ``shear-induced phase" and in the following we will refer to SIS as the reversible structures (or ``phase'') induced at larger scale in regime IV.

\section{Stress induced by gravity and development of shear-induced structures }\label{sec:SIS}
\begin{figure}
\centering
\includegraphics[scale=1]{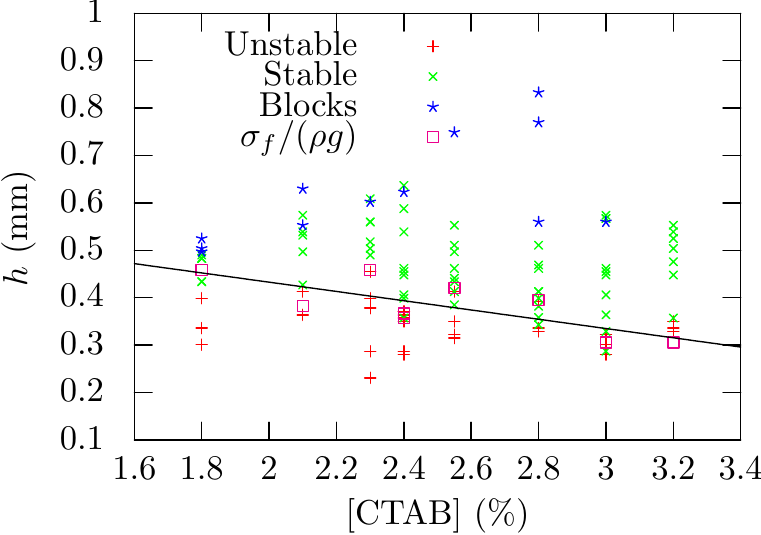}              
 \caption{Film morphologies for different film thicknesses and various surfactant concentrations ($[\textrm{NaSal}]=0.32$wt.\%). 
 The black line is a fit of film thicknesses $h_f=\sigma_f/(\rho g)$ ($\square$) related to stress plateau values. It separates unstable and stable flow regimes.}\label{fig:lindiagram}
\end{figure}

As described in section \ref{sec:flow_regime}, the film morphology depends on the film thickness $h$.
Our working hypothesis is that the transition between unstable to stable morphologies along the fibre might be linked to the presence of the SIS. 
To clarify and to go further on that point, we will focus on solutions in domain $\cal{C}$. 
This choice is motivated by the fact that solutions which belong to this domain present all three morphologies.
We first study the influence of the CTAB concentration.
We use a set of chemical solutions for which CTAB concentration is in the range $[1.8,3.2]$wt.\% keeping $[\textrm{NaSal}]=0.32$wt.\% constant.
For each solution, we keep increasing the flow rate.
The results are reported in figure \ref{fig:lindiagram} where we observe successively unstable, stable and block morphologies as the film thickness increases.

To explain this transition between stable and unstable morphologies, and since the rheology is controlled by the applied shear stress, we compare the stress plateau $\sigma_f$
extracted from the rheological curves to the gravitational stress $\sigma_g$ applied on the film. This stress $ \sigma_g = \rho g h$ ranges between 2 and 8 Pa in our experiments
and is then comparable to the value of the stress plateau $\sigma_f$. We define the film thickness $h_f$ related to the stress plateau $\sigma_f$ as $h_f = \sigma_f/(\rho g)$.

Figure \ref{fig:lindiagram} illustrates the succession of film morphologies as a function of the film thickness $h$ for different surfactant concentrations and a fixed salt concentration. We also report, for each sample, the value $h_f = \sigma_f/(\rho g)$ gathered from the rheological experiments. Remarkably, 
the unstable-stable transition is found to occur as $h\simeq h_f$.
Thus, we can conclude that the condition $\sigma_g>\sigma_f$ is necessary to stabilize the film. 
This suggests that the stabilization of the film is connected to the presence of the SIS as $\sigma_f$ is the characteristic stress value for the development of these structures. 

Consequently, the SIS are able to be formed, before the Rayleigh-Plateau instability acts, if $\sigma > \sigma_f$. This fixes the first condition to stabilize the film.
As we will see in the next section, this condition is necessary but not sufficient.

\section{Effect of the SIS elasticity on the instability growth rate}\label{sec:elasticity}

\begin{figure}
\centering               
\includegraphics[scale=1]{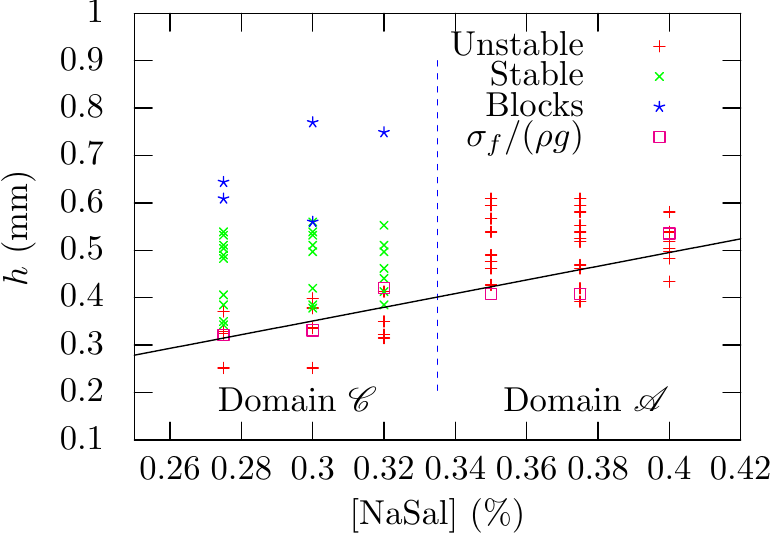}
 \caption{Film morphologies for different film thicknesses and various salt concentrations ($[\textrm{CTAB}]= 2.55$wt.\%).  
 The black line is a fit of film thicknesses $h_f=\sigma_f/(\rho g)$ related to stress plateau values.}\label{fig:lindiagram2}
\end{figure}

In this section, we study a second set of chemical solutions in order to complete our understanding.
The question we attempt to answer is: why do films become suddenly always unstable if the concentrations are slightly changed near the border between domains $\cal{A}$ and $\cal{C}$?
To cross these two domains, we work with a constant CTAB concentration ($[\textrm{CTAB}]=2.55$wt.\%) while NaSal concentrations vary in the range $[0.25,0.4]$wt.\%. 
Figure~\ref{fig:lindiagram2} displays the film morphology as a function of the thickness for different salt concentrations covering domains $\cal{A}$ and $\cal{C}$ at fixed surfactant concentration. Based on rheological data, the thickness of the film $\sigma_f/(\rho g)$ beyond which the SIS are likely to develop is also reported.
A good agreement is again observed in domain $\cal{C}$ between the expected thickness computed from rheological results and the effective thickness above which the films are stable.
However, there is a discrepancy for solutions in domain $\cal{A}$ since films thicker than $h_f$ are unstable. 
According to our analysis, the SIS still develop but they do not stabilize the film.

To understand these observations, we focus on two effects: the surface tension which is responsible for the instability and the bulk elasticity which tends to slow down the  film deformations.

It is worth noting that the constitutive relation describing the SIS is still unknown.
Since SIS present strong elastic properties, we attempt a description of the SIS by a simple Kelvin-Voigt model allowing analytic calculations \cite{Safran1993,Closa2012a}.
The Kelvin-Voigt model takes into account the viscous dissipation at short time scales and elasticity at long time scales.
The stress tensor $\sigma_{ij}$ is function of two parameters: viscosity $\eta_0$ and elastic modulus $G_0$:

\begin{equation}
 \sigma_{ij}(t) = G_0 \epsilon_{ij} + \eta_{0}  \dot\epsilon_{ij}\label{eq:kv}
\end{equation}
where $\epsilon_{ij}$ is the strain and the dot represents the time derivative.

To obtain the dispersion relation for this constitutive equation, we perform a linear stability calculation.
The Fourier transform of the expression (\ref{eq:kv}) reduces the equation to

\begin{equation}
 \sigma_{ij}(\omega) = \eta(\omega)  \dot\epsilon_{ij}(\omega)
\end{equation}

where $\omega$ is a complex frequency and with $\eta(\omega) = \frac{i G_0}{\omega} + \eta_{0}$.

Experimentally, we observe that the film thicknesses $h(z)$ and the fibre radius $R$ have the same order of magnitude.
Under this consideration, we can calculate the dispersion relation in our geometry\cite{Duprat2009b}.

In the cylindrical reference frame $(r,\theta, z)$ (See figure \ref{fig:setup}(a)), the $z$-component of the momentum balance, taking into account the lubrication approximation, becomes:
\begin{equation}
 \rho g - \partial_z p + \frac{\eta}{r}\partial_r(r\partial_r u)= 0
\end{equation}
where $u$ is the fluid velocity along the fibre ($z$ direction).
The boundary conditions are: no-slip on the fibre ($u(r=R)=0$) and zero tangential stress at the liquid-air interface ($\partial_r u (r=R+h)=0$).
The pressure gradient is caused by surface curvatures and it is expressed from Laplace's law: $\partial_z p = -\gamma \left( \frac{\partial_z h}{(R+h)^2} + \partial_{zzz}h\right)$ (assuming $\partial_z h \ll 1$). 
Solving this equation, the fluid velocity profile is $u(r) = \frac{\partial_z p- \rho g}{4\eta} \left[ (r^2-R^2) - 2(R+h)^2 \ln\left(\frac{r}{R}\right)\right]$.

Incompressibility leads to:
\begin{equation}
 \frac{\partial h}{\partial t} + \frac{\partial q}{\partial z} =0
\end{equation}
where the flow rate per unit length is $q=\frac{1}{2\pi(R+h)} \int_{R}^{R+h} u(r) 2\pi r dr$.

To conduct a linear stability analysis, we develop the film thickness as $h=h_0 + h_1 e^{i(kz-\omega t)}$ where $k$ is the real wave number.
Thus, we derive the following dispersion relation

\begin{equation}
\omega = k \frac{\rho  g R^2  \psi(\alpha) }{16 \eta(\omega)}  - i \frac{\gamma R h_0^3 \phi(\alpha)}{3\eta(\omega) (R+h_0)}  \left( k^4 - \frac{k^2}{(R+h_0)^2} \right)\label{eq:disp}
\end{equation}
with $\alpha=h_0/R$. 
Two dimensionless functions reflect the geometry:\\ $ \psi(x) = \frac{-x(2+x)(6+5x(2+x)) + 12(1+x)^4 \ln(1+x) }{(1+x)^2}$ and\\ $ \phi(x) = \frac{3(4(x+1)^4 \ln(x+1)   -x(x+2)(3x(x+2)+2))}{16x^3}\label{eq:phi_newtonien} $

\begin{figure}
\begin{center}
 \includegraphics{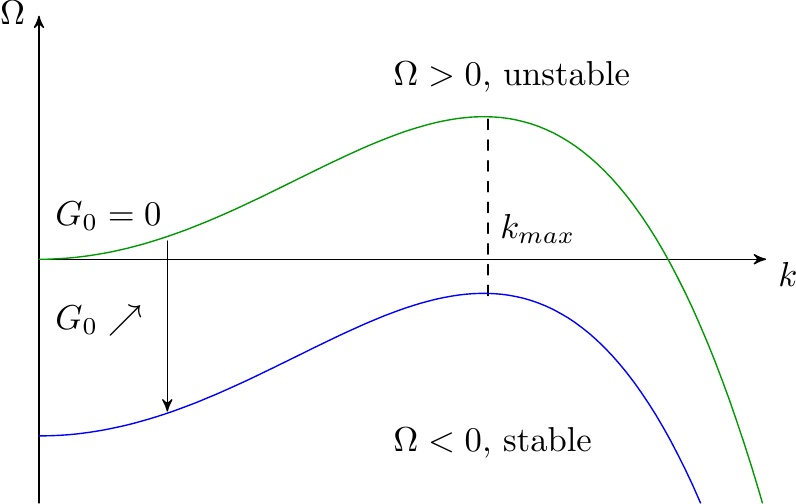}
\end{center}
 \caption{
 Growth rate of the instability $\Omega$ vs the wave number $k$ from the Kelvin-Voigt model illustrating the stabilizing effect of the bulk elasticity $G_0$.
 The case $G_0=0$ (in green) is the marginal stability curve for Newtonian fluids (viscosity $\eta_0$). 
 As $G_0$ increases, the marginal curves (in blue) is shifted to the negative growth rate domain.
 }\label{fig:stabilisation}
\end{figure}

The imaginary part of $\omega$ corresponds to the instability growth rate noted $\Omega(k)$:

\begin{equation}
 \Omega(k) = \frac{-\gamma R h_0^3 \phi(\alpha)}{3\eta_0 (R+h_0)}  \left( k^4 - \frac{k^2}{(R+h_0)^2} \right) - \frac{G_0}{\eta_0}\label{eq:disp2}
\end{equation}
For $\Omega(k)<0$, the system is stable, whereas for $\Omega(k)>0$ the system is unstable.
If $G_0=0$, the solution for Newtonian fluids is recovered.
The evolution of the growth rate with $k$ is plotted in figure \ref{fig:stabilisation}. 

We observe that the bulk elasticity ($G_0$) plays a stabilizing effect on the film as illustrated in figure \ref{fig:stabilisation}: the marginal stability curve is shifted to the negative $\Omega$.
If $\textrm{max}(\Omega) = \Omega(k_{max})<0$ (blue curve in figure \ref{fig:stabilisation}), all modes are damped by the elasticity resulting in a stable film.

Note that, in this model, the wave number for the maximum of the growth rate is $k_{max} = \frac{1}{\sqrt{2}(R+h_0)}$ and is independent of the elastic modulus.

Solving the condition $\Omega(k_{max}) = 0$ defines a critical elastic modulus $G_0^c$:
\begin{equation}
G_0^c(h) = \frac{\gamma h^3 R \phi(h/R)}{12(R+h)^5}\label{eq:critical_modulus}
\end{equation}

Experimentally, we can estimate $G_0^c$  from the border between domains $\cal{A}$ and $\cal{C}$. 
From the data presented in figure \ref{fig:lindiagram2}, we can evaluate for $[\textrm{CTAB}]\simeq2.55 \textrm{wt.}\%\simeq 70$mM, $[\textrm{NaSal}]\simeq0.33\textrm{wt.}\%\simeq20$mM: 
$G_0 \simeq G_0^c(h_f) = 1.0$ Pa (with $h=h_f=0.4$ mm).

Thus the bulk elasticity of the SIS has to be sufficient to shift the curve of the growth rate to the negative values,  in order to stabilize the film.
To summarize, we have shown that the elastic shear induced structures are developed in our flow if $\rho g h > \sigma_f$.
If this condition is fulfilled, we expect a stable film only if the bulk elasticity is greater than $G_0^c$.
As the result, we can explain the apparent discrepancy of the unstable region in figure \ref{fig:lindiagram2}. 
From the transient zone to the unstable one (i.e. increasing salt concentration), the bulk elasticity decreases such as $G_0<G_0^c$.

The value $G_0^c=1$ Pa can be compared to oscillatory shear flow tests. 
As suggested by the flow curve in figure \ref{fig:rheology}, the linear regime could not be reached with our rheometer. 
The first test consists in the measurement of the elastic modulus versus time for a shear stress amplitude of $5 \times 10^{-4}$ Pa at a frequency $f=1$Hz. 
The elastic modulus increases in time from $\sim 10^{-2}$ Pa to a few Pascal ($[\textrm{CTAB}]=2.55$ wt.\% and $[\textrm{NaSal}]=0.32$ wt.\%).
In a second test, a constant shear stress $\sigma>\sigma_f$ is applied to the sample followed by an oscillating stress of small amplitude. 
Even if the SIS has time to partially relax, such experiments can provide a reasonnable order of magnitude of the moduli ($G'$ and $G"$) of the SIS at high angular velocities. 
We found a value for the elastic modulus around $1.5$ Pa, consistent with the value $G_0^c=1$ Pa gathered from the flow along the fibre and considering that the viscoelastic properties of the SIS can be modelled by a Kelvin-Voigt model. 
Our results suggest that the elastic modulus of the solutions under shear is larger than the one for the solution at rest and that the value required to stabilize the flow could be deduced from the Kelvin-Voigt model.

\section{Conclusion}

In this paper, we studied the flow of CTAB/NaSal solutions on a vertical fibre. 
Structure modifications occuring from the chemical composition and the shear flow provides a large variety of flowing regimes. The flow on the fiber can actually be unstable and similar to the flow of newtonian fluids: the capillary driven Rayleigh-Plateau instability produces drops sliding on the fibre.
For another set of solutions, the film can stay uniform along the fiber provided the film thickness satisfies $h>h_f=\sigma_f/(\rho g)$. 
From rheo-optical measurements, we found that the condition for stabilizing the flow is that the characteristic stress on the film should be larger than a critical stress $\sigma_f$. This critical stress has been identified as the onset of emergence of micronic structures induced by the shear-flow, i.e., shear-induced structures (SIS). 
A last flowing regime is observed for high film thicknesses and in a given range of concentrations: in this last regime,  the film breaks in a series of gel-like blocks sliding along the fibre. 

Our analysis suggests that the bulk elasticity of the SIS is responsible for the decrease of the instability growth rate until inhibited the Rayleigh-Plateau instability for a negative growth rate. Two conditions are required to prevent the destabilization of the film: the presence of the SIS and a sufficient elasticity of theses structures to inhibit the instability driven by the surface tension. A linear analysis of stability assuming a Kelvin-Voigt model for SIS,  evidences a stable flow provided the elastic modulus of the SIS is higher than a critical value. By comparing with experimental results on systems which can  go through an unstable to stable regime, we can estimate the value of the elastic modulus value of these SIS. In the future, the inspection of the micronic SIS can provide a better understanding of the micellar structure inducing the bulk elasticity.

\section{Acknowledgment}

The authors thank Triangle de la Physique for the rheometer (Anton Paar, MCR 501) and F\'ed\'eration Paris VI for the high-speed camera.
Also, thanks to J\'er\^ome Delacotte, Christophe Clanet and Marina Moreno Luna for discussions.

\bibliography{micelle}

\begin{thebibliography}{10}

\bibitem{Berret2001}
J.-F Berret, R.~Gamez-Corrales, Y.~S\'er\'ero, F.~Molino, and P.~Lindner.
\newblock Shear-induced micellar growth in dilute surfactant solutions.
\newblock {\em Europhys. Lett.}, 54:605--611, 2001.

\bibitem{Berret2006}
J.F Berret.
\newblock {\em Rheology of wormlike micelles: equilibrium properties and shear
  banding transitions}, pages 667--720.
\newblock Springer, 2006.

\bibitem{Bhardwaj2007}
A.~Bhardwaj, E.~Miller, and J.~Rothstein.
\newblock Filament stretching and capillary breakup extensional rheometry
  measurements of viscoelastic wormlike micelle solutions.
\newblock {\em Journal of Rheology}, 51:693--719, 2007.

\bibitem{Bhat2010}
P.P Bhat, S.~Appathurai, M.T Harris, M.~Pasquali, G.H McKinley, and O.A
  Basaran.
\newblock Formation of beads-on-a-string structures during break-up of
  viscoelastic filaments.
\newblock {\em Nature Physics}, 6:625--631, 2010.

\bibitem{Boltenhagen1997}
P.~Boltenhagen, Y.~Hu, E.F. Matthys, and D.J. Pine.
\newblock Inhomogeneous structure formation and shear-thickening in worm-like
  micellar solutions.
\newblock {\em Europhys. Lett.}, 38:389 -- 394, 1997.

\bibitem{Boulogne2012}
F.~Boulogne, L.~Pauchard, and F.~Giorgiutti-Dauphin\'e.
\newblock Instability and morphology of polymer solutions coating a fiber.
\newblock {\em Journal of Fluid Mechanics}, 704:232, 2012.

\bibitem{Boys1959}
C.V Boys.
\newblock {\em Soap Bubbles: Their Colors and Forces Which Mold Them}.
\newblock Thomas Y. Crowell Company, 1959.

\bibitem{Cates2006}
M.~E. Cates and S.~M. Fielding.
\newblock Rheology of giant micelles.
\newblock {\em Advances in Physics}, 55:799--879, 2006.

\bibitem{Clasen2006}
C.~Clasen, J.~Eggers, M.A Fontelos, J.~LI, and G.H McKinley.
\newblock The beads-on-string structure of viscoelastic threads.
\newblock {\em Journal of Fluid Mechanics}, 556:283--308, 2006.

\bibitem{Closa2012a}
F.~Closa, F.~Ziebert, and E.~Rapha\"el.
\newblock Effects of in-plane elastic stress and normal external stress on
  viscoelastic thin film stability.
\newblock {\em Mathematical Modelling of Natural Phenomena}, 7:6--19, 2012.

\bibitem{Cooper-White2002}
J.~J. Cooper-White, R.~C. Crooks, and D.~V. Boger.
\newblock A drop impact study of worm-like viscoelastic surfactant solutions.
\newblock {\em Colloids and Surfaces A: Physicochemical and Engineering
  Aspects}, 210:105--123, 2002.

\bibitem{Derjaguin1943}
B.~Derjaguin.
\newblock On the thickness of the liquid film adhering to the walls of a vessel
  after emptying.
\newblock {\em Acta Physicochim. URSS}, 20:349--352, 1943.

\bibitem{Duprat2009b}
C.~Duprat, C.~Ruyer-Quil, and F.~Giorgiutti-Dauphin\'e.
\newblock Spatial evolution of a film flowing down a fiber.
\newblock {\em Physics of Fluids}, 21:042109, 2009.

\bibitem{Duprat2007}
C.~Duprat, C.~Ruyer-Quil, S.~Kalliadasis, and F.~Giorgiutti-Dauphin\'e.
\newblock Absolute and convective instabilities of a viscous film flowing down
  a vertical fiber.
\newblock {\em Phys. Rev. Lett.}, 98:244502, 2007.

\bibitem{Eggers2008}
J.~Eggers and E.~Villermaux.
\newblock Physics of liquid jets.
\newblock {\em Reports on Progress in Physics}, 71:036601, 2008.

\bibitem{Fardin2009}
M.~A. Fardin, B.~Lasne, O.~Cardoso, G.~Gr\'egoire, M.~Argentina, J.~P.
  Decruppe, and S.~Lerouge.
\newblock Taylor-like vortices in shear-banding flow of giant micelles.
\newblock {\em Phys. Rev. Lett.}, 103:028302, 2009.

\bibitem{Hartmann1997a}
V.~Hartmann and R.~Cressely.
\newblock Influence of sodium salicylate on the rheological behaviour of an
  aqueous ctab solution.
\newblock {\em Colloids and Surfaces A: Physicochemical and Engineering
  Aspects}, 121:151--162, 1997.

\bibitem{Hartmann1997}
V.~Hartmann and R.~Cressely.
\newblock Simple salts effects on the characteristics of the shear thickening
  exhibited by an aqueous micellar solution of ctab/nasal.
\newblock {\em Europhys. Lett.}, 40:691--696, 1997.

\bibitem{Hu1998a}
Y.~T. Hu, P.~Boltenhagen, E.~Matthys, and D.~J. Pine.
\newblock Shear thickening in low-concentration solutions of wormlike micelles.
  ii. slip, fracture, and stability of the shear-induced phase.
\newblock {\em Journal of Rheology}, 42:1209--1226, 1998.

\bibitem{Hu1998}
Y.~T. Hu, P.~Boltenhagen, and D.~J. Pine.
\newblock Shear thickening in low-concentration solutions of wormlike micelles.
  i. direct visualization of transient behavior and phase transitions.
\newblock {\em Journal of Rheology}, 42:1185--1208, 1998.

\bibitem{Israelachvili2011}
J.~N. Israelachvili.
\newblock {\em Intermolecular and Surface Forces}.
\newblock Academic Press, 2011.

\bibitem{Jayaraman2003}
A.~Jayaraman and A.~Belmonte.
\newblock Oscillations of a solid sphere falling through a wormlike micellar
  fluid.
\newblock {\em Phys. Rev. E}, 67:065301, 2003.

\bibitem{Landau1942}
L.~Landau and B.~Levich.
\newblock Dragging of a liquid by a moving plate.
\newblock {\em Acta Physicochim. URSS}, 17:42--54, 1942.

\bibitem{Lerouge2010}
S.~Lerouge and J.-F Berret.
\newblock Shear-induced transitions and instabilities in surfactant wormlike
  micelles.
\newblock {\em Advances in Polymer Science}, 230:1--71, 2010.

\bibitem{Lerouge2008}
S.~Lerouge, M.A Fardin, M.~Argentina, G.~Gr\'egoire, and O.~Cardoso.
\newblock Interface dynamics in shear-banding flow of giant micelles.
\newblock {\em Soft Matter}, 4:1808--1819, 2008.

\bibitem{Lin1994}
Z.~Lin, J.~J. Cai, L.~E. Scriven, and H.~T. Davis.
\newblock Spherical-to-wormlike micelle transition in ctab solutions.
\newblock {\em The Journal of Physical Chemistry}, 98:5984--5993, 1994.

\bibitem{Oda2000}
R.~Oda, V.~Weber, P.~Lindner, D.J Pine, E.~Mendes, and F.~Schloesser.
\newblock Time-resolved small-angle neutron scattering study of
  shear-thickening surfactant solutions after the cessation of flow.
\newblock {\em Langmuir}, 14:4859--4863, 2000.

\bibitem{Ohlendorf1986}
D.~Ohlendorf, W.~Interthal, and H.~Hoffmann.
\newblock Surfactant systems for drag reduction: Physico-chemical properties
  and rheological behaviour.
\newblock {\em Rheol. Acta}, 25(5):468--486, 1986.

\bibitem{Protzl1997}
B.~Pr\"otzl and J.~Springer.
\newblock Light scattering experiments on shear induced structures of micellar
  solutions.
\newblock {\em J. Colloid Interface Sci.}, 190:327--333, 1997.

\bibitem{Quere1990}
D.~Qu\'{e}r\'{e}.
\newblock Thin films flowing on vertical fibers.
\newblock {\em Europhysics Letters}, 13:721, 1990.

\bibitem{Quere1999}
D.~Qu\'{e}r\'{e}.
\newblock Fluid coating on a fiber.
\newblock {\em Annual Review of Fluid Mechanics}, 31:347--384, 1999.

\bibitem{Rao1987}
URK Rao, C.~Manohar, B.~S. Valaulikar, and R.~M. Iyer.
\newblock Micellar chain model for the origin of the visoelasticity in dilute
  surfactant solutions.
\newblock {\em The Journal of Physical Chemistry}, 91:3286--3291, 1987.

\bibitem{Rayleigh1878}
L.~Rayleigh.
\newblock On the instability of jets.
\newblock {\em Proceedings of the London Mathematical Society}, s1-10:4--13,
  1878.

\bibitem{Safran1993}
S.~A. Safran and J.~Klein.
\newblock Surface of viscoelastic thin films.
\newblock {\em J. Phys. II France}, 3:749--757, 1993.

\bibitem{schubert2004}
Beth~A. Schubert, Norman~J. Wagner, Eric~W. Kaler, and Srinivasa~R. Raghavan.
\newblock Shear-induced phase separation in solutions of wormlike micelles.
\newblock {\em Langmuir}, 20(9):3564--3573, 2004.

\bibitem{Smolka2003}
L.~B. Smolka and A.~Belmonte.
\newblock Drop pinch-off and filament dynamics of wormlike micellar fluids.
\newblock {\em Journal of Non-Newtonian Fluid Mechanics}, 115:1--25, 2003.

\bibitem{Vasudevan2010}
Mukund Vasudevan, Eric Buse, Donglai Lu, Hare Krishna, Ramki Kalyanaraman,
  Amy~Q Shen, Bamin Khomami, and Radhakrishna Sureshkumar.
\newblock Irreversible nanogel formation in surfactant solutions by microporous
  flow.
\newblock {\em Nature Materials}, 9(5):436--441, 2010.

\end{thebibliography}
\bibliographystyle{plain}

\end{document}